\documentclass{ws-procs9x6}
\usepackage{epsfig}

\begin{document}

\centerline{\bf ETA PHOTOPRODUCTION OFF THE NEUTRON AT GRAAL }

\vspace{0.5  cm}
\begin{center}

V.Kuznetsov$^1$\footnote{E-mail \uppercase{S}lava@cpc.inr.ac.ru, \uppercase{S}lavaK@jlab.org},
O. Bartalini$^2$, V. Bellini$^3$, M. Castoldi$^4$, A. D'Angelo$^2$,
J-P. Didelez$^5$, R. Di Salvo$^2$, A. Fantini$^2$, D. Franco$^2$, G. Gervino$^6$, F. Ghio$^7$,
B. Girolami$^7$, A. Giusa$^3$, M. Guidal$^5$, E. Hourany$^5$, R. Kunne$^5$,
A. Lapik$^1$, P. Levi Sandri$^8$, D. Moricciani$^2$,
L. Nicoletti$^3$, C. Randieri$^3$, N. Rudnev$^{9}$, G.
Russo$^3$, C. Schaerf$^2$, M.-L. Sperduto$^3$, M.-C. Sutera$^3$, A. Turinge$^{10}$.

\vspace{0.5cm}

{\it $^1$Institute for Nuclear Research, 117312 Moscow, Russia}\\
{\it $^2$INFN sezione di Roma II and Universit\`a di Roma "Tor Vergata",
00133 Roma, Italy}\\
{\it $^3$INFN Laboratori Nazionali del Sud and Universit\`a di Catania,
95123 Catania,Italy}\\
{\it $^4$INFN Genova and Universit\`a di Genova, 16146 Genova, Italy }\\
{\it $^5$IN2P3, Institut de Physique Nucl\'eaire, 91406 Orsay, France}\\
{\it $^6$INFN sezione di Torino  and Universit\`a di Torino, 10125 Torino,Italy}\\
{\it $^7$INFN sezione Sanit\`a and Istituto Superiore di Sanit\`a,00161 Roma, Italy}\\
{\it $^8$INFN Laboratori Nazionali di Frascati, 00044 Frascati, Italy}\\
{\it$^{9}$Institute of Theoretical and Experimental Physics, Moscow, Russia}\\
{\it $^{10}$RRC "Kurchatov Institute", Moscow, Russia}

\end{center}
\vspace{0.5 cm}

\abstracts{\small
  The $\gamma n \rightarrow \eta n$ quasi-free cross section 
  reveals a resonant structure at $W\sim 1.675$ GeV. This structure may
  be a manifestation of a baryon resonance. {\it A priori} its
  propreties, the possibly narrow width and the strong photocoupling
  to the neutron, look surprising. This structure may also signal the
  existence of a narrow state. 
}

Meson photoproduction off the neutron offers an attractive tool to study certain
baryon resonances. A single-quark
transition model\cite{mok} predicts only weak photoexcitation 
of the $D_{15}(1675)$ resonance from the proton target. 
Photocouplings to the neutron calculated in the framework 
of this approach, are not small. An isobar model for 
$\eta$ photo- and electroproduction $\eta$-MAID\cite{maid}
also suggests significant contribution of the $D_{15}(1675)$ to $\eta$
photoproduction on the neutron.

Possible photoexcitation of the non-strange pentaquark is of high interest
as well. A benchmark signature of this particle (if it exists) 
is its photoproduction off the nucleon. The chiral soliton model
predicts that photoexcitation of the non-strange pentaquark has to be 
suppressed on the proton and should mainly occur on the neutron~\cite{max}. 
Estimates of the chiral soliton approach ranges its mass to $1.65 - 1.7$
GeV\cite{dia1}. Modified parial wave analysis of $\pi N$ scattering suggests two possible candidates,
at 1.68 and/or 1.73 GeV, with the total width about 10 MeV.
Among various reactions, $\eta$ photoproduction has been considered as particularly 
sensitive to the signal of this particle~\cite{max,dia1,str,jafw}. 

Up to now, $\eta$ photoproduction off the neutron was explored 
mostly in the region of the $S_{11}(1535)$ resonance from threshold 
up to W$\sim$1.6~GeV~\cite{inc}. The ratio of the $\gamma n \rightarrow \eta 
n$/$\gamma p\rightarrow\eta p$ cross sections was extracted and found  
almost constant  near $\sim$0.67.
At higher energies, the GRAAL Collaboration reported the sharp rise of
this ratio\cite{nstar2002} and the evidence for a resonant structure
in the cross section on the neutron at W$\sim$1.675~GeV~\cite{gr1}. 
Recently the CB/TAPS Collaboration reported similar observation\cite{kru1}.

\begin{figure}[ht]
\vspace {-0.2cm}
\centerline{
\psfig{file=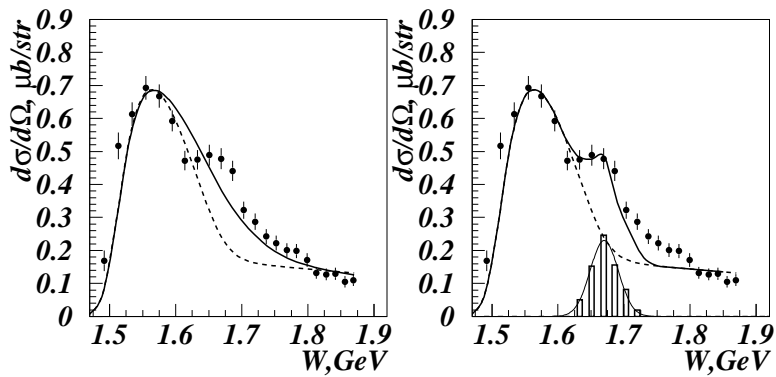,width=4.0in,clip=0,silent=,angle=0}
}
\caption{$\gamma n \rightarrow \eta n$ 
         quasi-free differential cross-section
         at $137^{\circ}$.
         Left panel:  Soild line is the  $\eta$-MAID
         prediction folded with Fermi motion. Dashed line is
         the same prediction without the $D_{15}(1675)$ resonance.
         Right panel: Dashed area shows the contribution of a narrow
         state. Solid line is the sum of the $\eta$-MAID cross section
         without $D_{15}(1675)$, folded with Fermi motion, and
         the narrow state. Dashed line is the same as in the left panel.
          \label{fig:cr2}}
\vspace*{-0.5cm}
\end{figure}

\begin{figure}[ht]
\vspace*{-0.2 cm}
\psfig{file=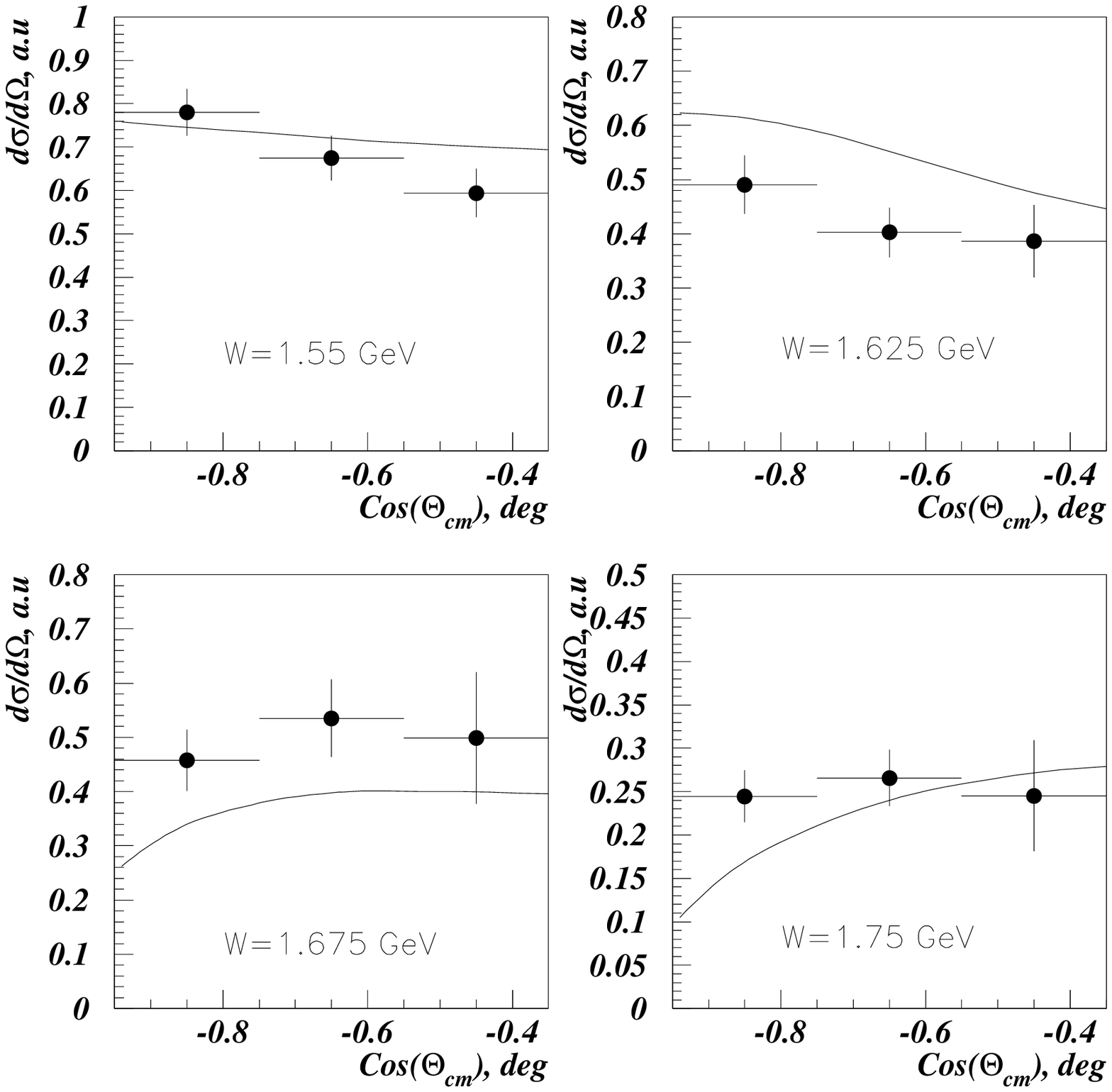,width=2.0in,clip=0,silent=0,angle=0}
\hspace{0.3cm}\psfig{file=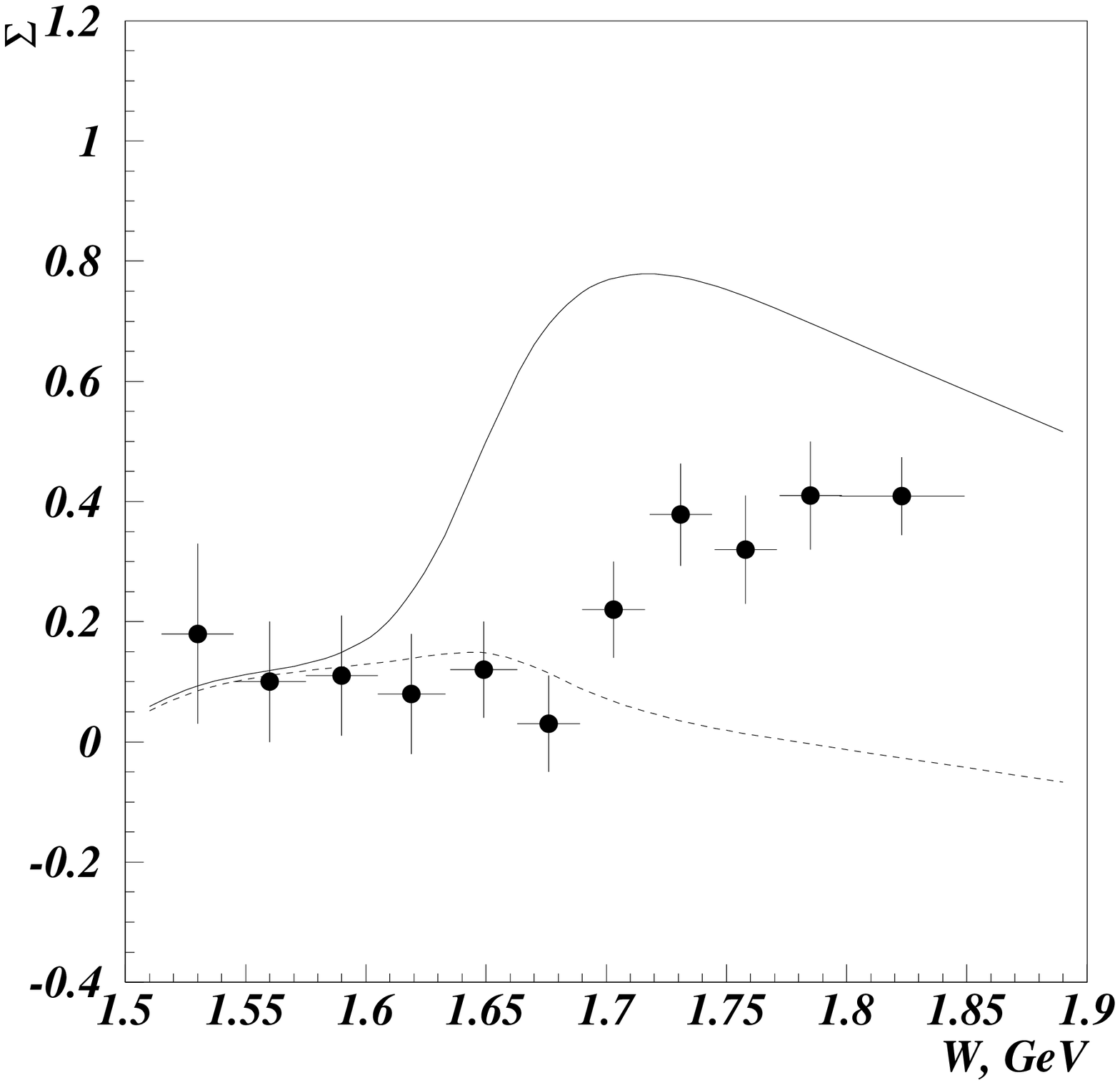,width=2.0in,clip=0,silent=0,angle=0}
\vspace*{0  cm}
\caption{Left panel: angular dependence of the $\gamma n \rightarrow \eta n$
  quasi-free cross section. Right panel: $\gamma n \rightarrow \eta n$
  beam asymmetry $\Sigma$ at 137$^{\circ}$. Solid lines are $\eta$-MAID predictions.
  Solid and dashes lines are $\eta$-MAID predictions with all resonances and
  without $D_{15}(1675)$ respectively.  
\label{fig:cr3}}
\vspace{-0.3cm}
\end{figure}

Quasi-free $\eta n$ differential photoproduction cross section 
is shown in Fig.~\ref{fig:cr2}. The cross section clearly reveals 
a resonant structure near $W\sim 1.675$ GeV.
We compared  this cross section with an isobar model for 
$\eta$ photo- and electroproduction $\eta - MAID$\cite{maid}.
The model includes 8 main resonances and suggests the dominance of the
$S_{11}(1535)$ and $D_{15}(1675)$ resonances in $\eta$ photoproduction
off the neutron below $W\sim 1.7 $ GeV\footnote{We refer to the
recent update of $\eta - MAID$ which includes
the corrected helicity amplitude $^{n}A_{\frac{1}{2}}$ 
of the $D_{15}(1675)$ resonance.}.   
The model predicts a bump-like structure near $W \sim 1.675$ GeV
in the total $\eta$ photoproduction cross section on the
neutron\cite{tia}. This structure is caused by the $D_{15}(1675)$ resonance. 
The $\eta$ - MAID differential cross section at $137^{\circ}$ is  
smooth (Fig.~\ref{fig:cr2}, left panel).  
$\eta - MAID$ reasonably reproduces the angular dependence of the cross section while predicts 
larger beam asymmetries above $W\sim 1.6$ GeV(Fig.~\ref{fig:cr3}). 

The PDG estimate for the Breit-Wigner width of the
$D_{15}(1675)$ resonance is $\Gamma \sim 150$ MeV\cite{pdg}. 
The structure observed in the quasi-free cross section
looks  more narrow. $\eta - MAID$ without 
the $D_{15}(1675)$ resonance fits the cross section 
in the region of the $S_{11}(1535)$ resonance below 
$W\sim 1.62$ GeV(Fig.~\ref{fig:cr2}). One may assume that 
above this region there is a contribution of an additional 
relatively narrow resonance. 

In the right panel of Fig.~\ref{fig:cr2}, 
a simulated narrow ($M=1675$ GeV, $\Gamma = 10$ MeV) state is shown. 
This state appears as a bump in the quasi-free cross section due to
Fermi motion of the target neutron bound in a deuteron target.
The sum of the $\eta - MAID$ without $D_{15}(1675)$
and the narrow state well reproduces the cross section 
up to $W \sim 1.7$ GeV. At higher energies, the increasing 
contribution of higher-lying
resonances is expected\cite{etap}.
Thus, the structure in the $\gamma n \rightarrow \eta n$ cross
section may signal the existence of a relatively narrow state.
If confirmed, such state coincide with the expectation of the
Chiral Soluton Model\cite{max,dia1} and modified PWA\cite{str} for the
non-strange pentaquark. On the other hand,
 the manifectation of one of
usual resonances is not ruled out. {\it Apriori} its properties,
the possibly narrow width and the strong photocoupling to the neutron,
look surprising. More data and detailed partial wave analysis are
needed to identify the nature of the observed structure. New programs to
study $\eta$ photorpoduction off the neutron with polarized targets,
which are now launched at modern photon factories such as the upgraded 
MamiC Facility (Mainz, Germany), aim to provide experimental data at new 
level of quality.

I wish to thank Bill Briscoe and Igor Strakovsky for assistance in 
preparation and delivery of this talk, Berndt Krusche for fruitful 
discussions, and Lothar Tiator for providing new $\eta$ - MAID 
predictions.


\end{document}